\documentclass[a4paper,10pt,twoside]{cpc-hepnp}

\usepackage{subfigure}
\usepackage{multicol}
\usepackage{units}
\usepackage{booktabs}
\usepackage{amsmath}
\usepackage{amsfonts,amssymb,bm,mathrsfs,bbm,amscd}
\usepackage{stfloats}
\usepackage{lastpage}
\usepackage{graphicx}
\usepackage{multirow}
\DeclareGraphicsExtensions{.pdf,.jpeg,.png}
\usepackage{epstopdf}
\usepackage{lineno}
\usepackage[english]{babel}
\usepackage[colorlinks=true,pdfpagemode=FullScreen,,setpagesize=off,pdfborder={0 0 0}]{hyperref}
\usepackage{overpic}
\usepackage{lineno}
\usepackage{color}
\usepackage{diagbox}
\usepackage{xurl}
\usepackage{caption}

\usepackage{soul} 
\usepackage{float}
\usepackage{ulem}
\usepackage[fontset=macnew]{ctex}

\begin{document}

\fancyhead[c]{\small Chinese Physics C~~~Vol. xx, No. x (2024) xxxxxx}
\fancyfoot[C]{\small 010201-\thepage}
\footnotetext[0]{Received xxxx Month xxxx}

\title{\boldmath Measurement of the integrated luminosity of the data
 collected at 3.773~GeV by BESIII from 2021 to 2024\thanks{
The BESIII Collaboration thanks the staff of BEPCII and the IHEP computing center 
for their strong support. This work is supported in part by National Key R$\&$D 
Program of China under Contracts Nos. 2020YFA0406400, 2020YFA0406300, 2023YFA1606000; 
National Natural Science Foundation of China (NSFC) under Contracts Nos. 123B2077, 12035009, 11635010, 
11735014, 11875054, 11935015, 11935016, 11935018, 11961141012,  12025502, 12035013, 
12061131003, 12192260, 12192261, 12192262, 12192263, 12192264, 12192265, 12221005, 
12225509, 12235017, 12361141819; the Chinese Academy of Sciences (CAS) Large-Scale 
Scientific Facility Program; the CAS Center for Excellence in Particle Physics (CCEPP); 
Joint Large-Scale Scientific Facility Funds of the NSFC and CAS under Contract Nos. U2032104, U1832207; 
the Excellent Youth Foundation of Henan Scientific Commitee under Contract No.~242300421044;
100 Talents Program of CAS; The Institute of Nuclear and Particle Physics (INPAC) and 
Shanghai Key Laboratory for Particle Physics and Cosmology; German Research Foundation DFG 
under Contracts Nos. 455635585, FOR5327, GRK 2149; Istituto Nazionale di Fisica Nucleare, 
Italy; Ministry of Development of Turkey under Contract No. DPT2006K-120470; National 
Research Foundation of Korea under Contract No. NRF-2022R1A2C1092335; National Science 
and Technology fund of Mongolia; National Science Research and Innovation Fund (NSRF) 
via the Program Management Unit for Human Resources $\&$ Institutional Development, 
Research and Innovation of Thailand under Contract No. B16F640076; Polish National 
Science Centre under Contract No. 2019/35/O/ST2/02907; The Swedish Research Council; 
U. S. Department of Energy under Contract No. DE-FG02-05ER41374.}
}
\maketitle
\begin{center} 
M.~Ablikim$^{1}$, M.~N.~Achasov$^{4,c}$, P.~Adlarson$^{75}$, O.~Afedulidis$^{3}$, X.~C.~Ai$^{80}$, R.~Aliberti$^{35}$, A.~Amoroso$^{74A,74C}$, Q.~An$^{71,58,a}$, Y.~Bai$^{57}$, O.~Bakina$^{36}$
% ...
\\
\vspace{0.2cm}
(BESIII Collaboration)\\
\vspace{0.2cm} {\it
$^{1}$ Institute of High Energy Physics, Beijing 100049, People's Republic of China\\
% ...
$^{3}$ Bochum  Ruhr-University, D-44780 Bochum, Germany\\
$^{4}$ Budker Institute of Nuclear Physics SB RAS (BINP), Novosibirsk 630090, Russia\\
% ...
$^{35}$ Johannes Gutenberg University of Mainz, Johann-Joachim-Becher-Weg 45, D-55099 Mainz, Germany\\
$^{36}$ Joint Institute for Nuclear Research, 141980 Dubna, Moscow region, Russia\\
% ...
$^{57}$ Southeast University, Nanjing 211100, People's Republic of China\\
$^{58}$ State Key Laboratory of Particle Detection and Electronics, Beijing 100049, Hefei 230026, People's Republic of China\\
% ...
$^{71}$ University of Science and Technology of China, Hefei 230026, People's Republic of China\\
% ...
$^{74}$ University of Turin and INFN, (A)University of Turin, I-10125, Turin, Italy; (B)University of Eastern Piedmont, I-15121, Alessandria, Italy; (C)INFN, I-10125, Turin, Italy\\
$^{75}$ Uppsala University, Box 516, SE-75120 Uppsala, Sweden\\
% ...
$^{80}$ Zhengzhou University, Zhengzhou 450001, People's Republic of China\\

\vspace{0.2cm}
$^{a}$ Deceased\\
% ...
$^{c}$ Also at the Novosibirsk State University, Novosibirsk, 630090, Russia\\
% ...

}

\end{center}

\begin{abstract}
  We present a measurement of the integrated luminosity of $e^+e^-$  collision
  data collected with the BESIII detector at the BEPCII collider at a
  center-of-mass energy of $E_{\rm cm} = 3.773$~GeV. The integrated
  luminosities of the data sets taken from December 2021 to June 2022, from
  November 2022 to June 2023, and from October 2023 to February 2024 are
  determined to be $4.995 \pm 0.019$~fb$^{-1}$, $8.157 \pm 0.031$~fb$^{-1}$,
  and $4.191 \pm 0.016$~fb$^{-1}$, respectively, by analyzing large angle
  Bhabha scattering events. The uncertainties are dominated by systematic
  effects and the statistical uncertainties are negligible. Our results provide
  essential input for future analyses and precision measurements.
\end{abstract}


\begin{multicols}{2}

\section{Introduction}
Luminosity plays a crucial role in quantifying the size of a dataset and is a
fundamental parameter for measuring various physics processes, particularly
cross sections. The number of events for the process $e^+e^-\to X$ ($X$ denotes
any possible final state) in $e^+e^-$ collision data can be expressed as
\begin{eqnarray}
  N_{e^+e^-\to X} = \mathcal{L} \times \sigma_{e^+e^-\to X} (E_{\rm cm}),
  \label{eq:chi_TOF}
\end{eqnarray}
where $\mathcal{L}$ denotes the integrated luminosity of the data set,
$\sigma_{e^+e^-\to X}$ is the cross section for the process $e^+e^-\to X$, and
$E_{\rm cm}$ is the center-of-mass energy. In principle, any process with a
known cross section can be used to determine luminosity. However, Quantum
Electrodynamics~(QED) processes are advantageous due to their high production
rates, simple final state topologies, and cross sections that are known with high
theoretical precision. 

Datasets at $E_{\rm cm} = 3.773$~GeV were collected by the BESIII detector at
the BEPCII collider from December 2021 to June 2022~(DATA~I), from November
2022 to June 2023~(DATA~II), and from October 2023 to February 2024~(DATA~III)
in order to systematically investigate the properties of $\psi(3770)$ and $D$
meson production and decays. Precisely determining the luminosity of this
dataset is important for various purposes. It is essential for measuring the
cross section of $e^+e^-\to \psi(3770) \to D\bar{D}$, calculating the
normalization factors in strong-phase measurements of $D^0$
decays~\cite{BESIII:2020hlg, BESIII:2020khq}, and reducing the systematic
uncertainty of analyses using the single tag method~\cite{BESIII:stm}.
Additionally, the luminosity is used to normalize the Monte Carlo~(MC) sample
size and to estimate continuous background in the $\psi(3686)$
dataset~\cite{BESIII:3686, BESIII:chicj}. This paper presents a measurement of
the integrated luminosity in DATA~I, DATA~II, and DATA~III using large angle
Bhabha scattering events. This dataset is currently the world's largest
collection of $e^+e^-$ collision data at the $\psi(3770)$ resonance peak.

\section{BESIII detector and Monte Carlo simulation}
The BESIII detector~\cite{Ablikim:2009aa} records energy-symmetric $e^+e^-$
collisions provided by the BEPCII storage ring~\cite{Yu:IPAC2016-TUYA01} in the
$E_{\rm cm}$ range from 2.0 to 4.95~GeV, with a peak luminosity of
$1.1 \times 10^{33}\;\text{cm}^{-2}\text{s}^{-1}$ achieved at
$E_{\rm cm} = 3.773\;\text{GeV}$ in 2023. BESIII has collected large data
samples in this energy region~\cite{Ablikim:2019hff, EcmsMea, EventFilter}. The
cylindrical core of the BESIII detector covers 93\% of the full solid angle and
consists of a helium-based multilayer drift chamber~(MDC), a plastic
scintillator time-of-flight system~(TOF), and a CsI(Tl) electromagnetic
calorimeter~(EMC), which are all enclosed in a superconducting solenoidal
magnet providing a 1.0~T magnetic field. The solenoid is supported by an
octagonal flux-return yoke with resistive plate counter muon identification
modules interleaved with steel. The charged-particle momentum resolution at
$1~{\rm GeV}/c$ is $0.5\%$, and the ${\rm d}E/{\rm d}x$ resolution is $6\%$
for electrons from Bhabha scattering. The EMC measures photon energies with a
resolution of $2.5\%$~($5\%$) at $1$~GeV in the barrel~(end cap) region. The
time resolution in the TOF barrel region is 68~ps, while that in the end cap
region is 60~ps~\cite{etof}.

Simulated data samples produced with a {\sc geant4}-based~\cite{geant4} Monte
Carlo~(MC) package, which includes the geometric description of the BESIII
detector and the detector response, are used to determine detection
efficiencies and to estimate backgrounds. The QED processes are simulated with
the Babayaga@NLO
generator~\cite{babayaganew, Balossini:2006wc, Balossini:2008xr, theoretical}. 
The width of the $\psi (3770)$, beam energy spread, initial state
radiation~(ISR), and final state radiation~(FSR) are considered in the
simulation. The configuration parameters of the Bhabha signal
MC~($e^+e^-\to(\gamma)e^+e^-$) are listed in Table~\ref{optionMC}. As for other
processes, the simulation models the beam energy spread and ISR in the $e^+e^-$
annihilations with the generator {\sc kkmc}~\cite{ref:kkmc}. The inclusive MC
sample is used to simulate all possible processes of $e^+e^-$ collision,
including the production of $D\bar{D}$ pairs (including quantum coherence for
the neutral $D$ channels), the non-$D\bar{D}$ decays of the $\psi(3770)$, the
ISR production of the $J/\psi$ and $\psi(3686)$ states, and the continuum
processes incorporated in {\sc kkmc}~\cite{ref:kkmc}. 

All particle decays are modelled with {\sc evtgen}~\cite{ref:evtgen} using
branching fractions either taken from the Particle Data Group~\cite{PDG}, when
available, or otherwise estimated with {\sc lundcharm}~\cite{ref:lundcharm}.
FSR from charged final state particles is incorporated using the {\sc photos}
package~\cite{photos}.

In the simulation above, the run-by-run based calibration is applied, taking
into consideration the beam energy fluctuations and other time-dependent
effects. The duration of each run is typically one hour. The calibration of
$E_{\rm cm}$, which takes advantage of the small uncertainty in the known $D$
mass $m_{D}$~\cite{PDG} and the excellent resolution of the MDC and EMC, is
carried out using the following equations:
\begin{eqnarray}
  E_{\rm cm}=2E_{D}\,, E^2_{D}=E_0^2+(m_{D}c^2)^2-(M^{\rm fit}_{\rm BC}c^2)^2.
  \label{eq:Ecm}
\end{eqnarray}
Here, $E_0$ is the uncalibrated beam energy, i.e., 3.773~GeV, and
$M^{\rm fit}_{\rm BC}$ denotes the fitted peak of the beam-constrained mass of
the $D$ mesons. This is calculated using
$M^{\rm fit}_{\rm BC}c^2=\sqrt{E_0^2-p_D^2c^2}$, where $p_D$ is the momentum of
the $D$ measured in the center-of-mass system of the $e^+e^-$ collision using
the decays $D^{0}\to K^-\pi^+$, $D^{0}\to K^-\pi^+\pi^+\pi^-$ and
$D^{+}\to K^-\pi^+\pi^+$. Essentially, Eq.~\ref{eq:Ecm} is equivalent to
$E^2_D=p^2c^2+m^2_Dc^4$. The distributions of $M_{\rm BC}$, instead of $p_D$,
are utilized because $M_{\rm BC}$ offers better resolution and the fit to
$M_{\rm BC}$ is more easily controlled. According to the above analysis, the
calibrated $E_{\rm cm}$ varies by 2-3~MeV around the expected 3.773~GeV with an
uncertainty of approximately $0.02$~MeV. Details of the fit and particle
reconstruction are introduced in Ref.~\cite{BESIII:2022ulv}.

\begin{center}
  \tabcaption{Configuration of the Babayaga@NLO generator used to simulate Bhabha events.} \footnotesize  
  \begin{tabular}{cc}
    \toprule
    \hline
    Parameter & Value \\
    \hline
    Center-of-mass energy            & Calibrated $E_{\rm cm}$\\
    Beam Energy Spread     & 0.97~MeV \\
    Minimum $\cos\theta$   & -0.83 \\ 
    Maximum $\cos\theta$   & 0.83\\ 
      \hline
      \bottomrule\label{optionMC}
    \end{tabular}
  \end{center}  

\section{Method}
The integrated luminosity of data is usually measured with three QED processes:
$e^{+}e^{-}\to (\gamma)e^{+}e^{-}$, $e^+e^-\to (\gamma)\gamma\gamma$ and
$e^+e^-\to (\gamma)\mu^+\mu^-$. The symbol ``($\gamma$)" represents the
possible presence of photon(s) resulting from ISR or FSR. Since the
uncertainties associated with reconstructing electrons are smaller than those
of photons and muons, and the statistical uncertainty of Bhabha events are
almost negligible, only Bhabha event is used to measure luminosity.
Furthormore, Bhabha events at small angles, i.e., in the direction of the
$e^+ e^-$ beam, predominantly interact with the end cap region of the detector.
This region exhibits more gaps and worse resolution compared to the barrel
region. The generator simulation at small angles is also less precise. Given
that large angle Bhabha events provide a negligible statistical uncertainty, we
exclusively utilize them to measure the integrated luminosity of the data set.
The large angle region is defined as $|\cos\theta|<0.83$, where $\theta$ is the
polar angle of the final state electron~(positron) relative to the beam
direction.

The integrated luminosity of data is determined by
\begin{eqnarray}
  \mathcal{L} = \sum_i\frac{N_i^{\rm obs}\times(1-\eta)}{\sigma_i\times\epsilon_i\times\epsilon^{\rm trig}}\,.
  \label{eq:cal_lum}
\end{eqnarray}
In this equation, the index $i$ indicates different runs, and the integrated
luminosity of the data set is obtained by summing over these runs. The
variables $N_i^{\rm obs}$, $\sigma_i$, $\epsilon_i$, $\eta$, and
$\epsilon^{\rm trig}$ represent the number of observed Bhabha-event candidates,
the production cross section of the Bhabha process, the detection efficiency,
the contamination rate, and the trigger efficiency for collecting events in the
on-line data acquisition system, respectively. The Babayaga@NLO
generator~\cite{babayaganew, Balossini:2006wc, Balossini:2008xr, theoretical}
is employed, using the $E_{\rm cm}$ calibrated on a run-by-run basis,
to calculate the cross section, generate signal MC events for the process
$e^{+}e^{-}\to (\gamma)e^{+}e^{-}$, and estimate the detection efficiency.

\section{Luminosity measurement}
\subsection{Event selection}
\label{sec:selection}
Candidate Bhabha events are required to have exactly two oppositely charged
tracks detected in the MDC. Each track is required to satisfy a distance of
closest approach to the interaction point of less than 10~cm along the $z$
axis~(the symmetry axis of the MDC) and less than 1~cm in the transverse plane.
Additionally, the each candidate track must lie within the polar angle region
$|\rm cos\theta|<0.8$ to ensure interaction with the barrel of the EMC.

To suppress the $e^+e^-\to (\gamma)\mu^+\mu^-$ background, the deposited
energy $E_{\rm EMC}$ in the EMC by each track, shown in Fig.~\ref{fig:E_EMC},
must fall within the range $1.0<E_{\rm EMC}<2.5$~GeV. Furthermore, the sum of
the momenta of the two tracks, shown in Fig.~\ref{fig:p}, is required to be
larger than $0.9\times E_{\rm cm}$ to suppress background events, which mainly
from processes involving a $J/\psi$ in the final state, e.g., 
$e^+e^-\to(\gamma)J/\psi$, $e^+e^-\to(\gamma)\psi(3686)\to(\gamma)J/\psi X$,
and $e^+e^-\to\psi(3770)\to(\gamma)J/\psi X$. To eliminate the background from
energetic cosmic rays, the momentum of each track must be less than
$E_{\rm cm}/2 + 0.30$~GeV. 

The two oppositely charged tracks in the candidate Bhabha scattering events are
produced to be back-to-back. However, due to the presence of a magnetic field,
their trajectories are bent, resulting in their respective shower clusters in
the $xy$-plane of the EMC not being back-to-back. To quantify this angular
difference, the variable $\delta\phi$ is defined as
$|\phi_1-\phi_2|-180^{\circ}$, where $\phi_1$ and $\phi_2$ are the azimuthal
angles of the two clusters in the EMC. The distribution of $\delta\phi$ is
illustrated in Fig.~\ref{fig:delphi} and a requirement of
$5^{\circ}<|\delta\phi|<40^{\circ}$ is imposed. This requirement effectively
eliminates the background from the $e^+e^-\to (\gamma)\gamma\gamma$ process.

\begin{center}
  \includegraphics[width=7cm,height=7cm]{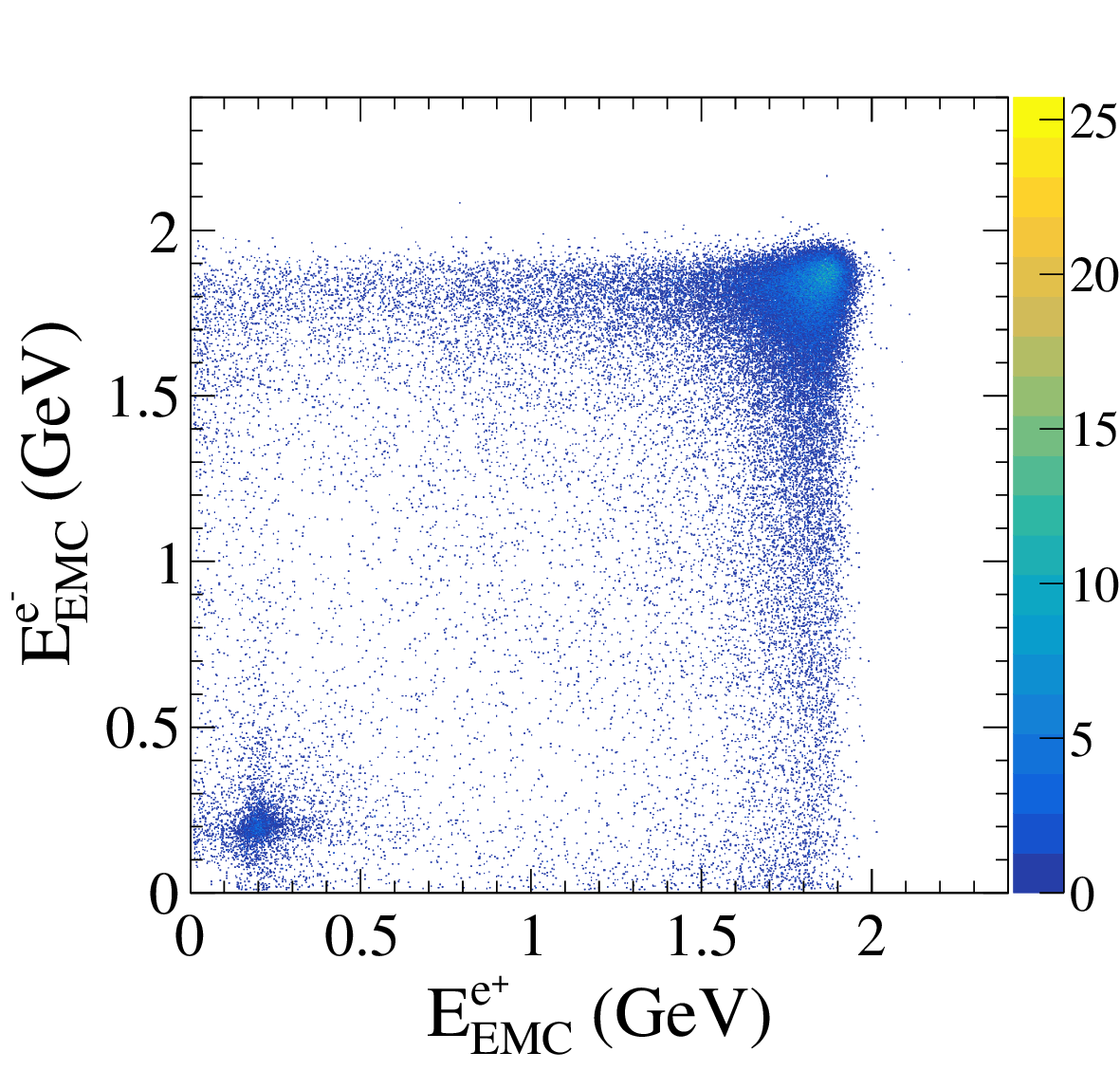}
  \figcaption{
    The distribution of $E^{e^+}_{\rm EMC}$ versus $E^{e^-}_{\rm EMC}$ from a
    subset of the data sample. The cluster concentrated at the upper right
    corner is the Bhabha signal and the small cluster at the lower left corner
    is the $e^+e^-\to (\gamma)\mu^+\mu^-$ background. The horizontal and
    vertical bands are caused by ISR and FSR in Bhabha and
    $e^+e^-\to (\gamma)\mu^+\mu^-$ events.}
  \label{fig:E_EMC}
\end{center}

\begin{center}
  \includegraphics[width=7cm]{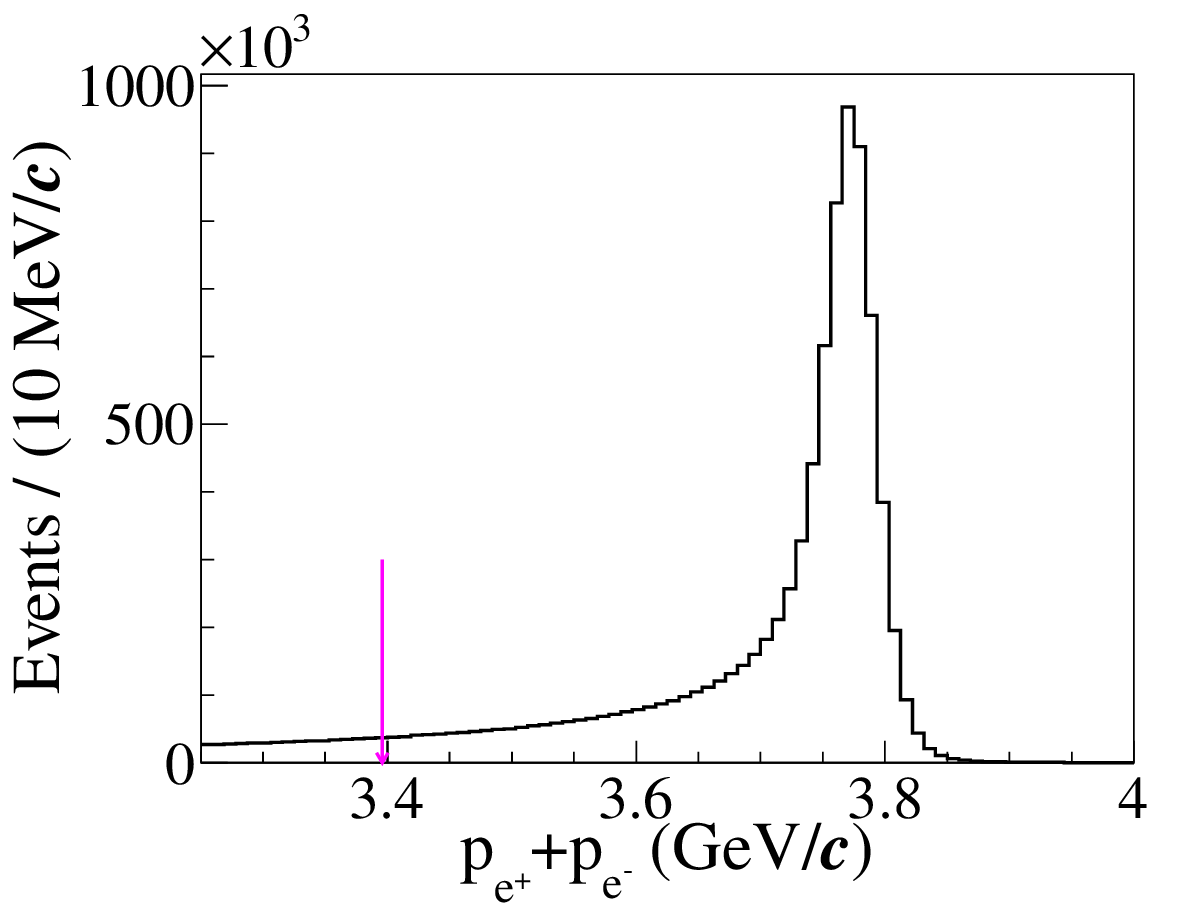}
  \figcaption{
    The distribution of $p_{e^+}+p_{e^-}$ from a subset of the data sample.
    The background is not visible due to the very low contamination rate of
    $\eta=3\times 10^{-4}$. The pink arrow indicates the
    $>0.9\times E_{\rm cm}$ requirement. 
  }
  \label{fig:p}
\end{center}

\begin{center}
  \includegraphics[width=7cm]{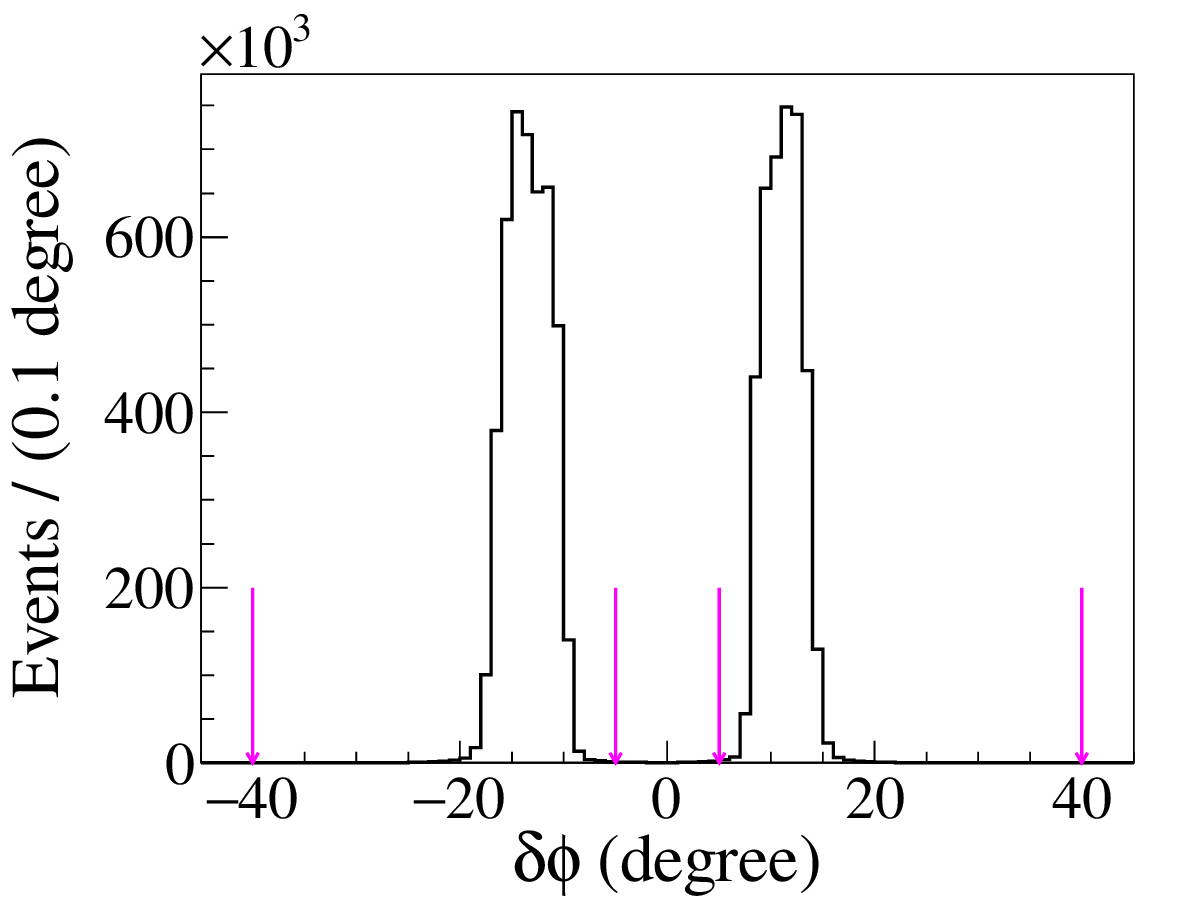}
  \figcaption{
    The distribution of $\delta \phi$ between the selected $e^+$ and $e^-$
    tracks from a subset of the data sample. The background is not visible due
    to the very low contamination rate of $\eta=3\times 10^{-4}$. The pink
    arrows indicate the $5^{\circ}<|\delta\phi|<40^{\circ}$ requirement.
  }
  \label{fig:delphi}
\end{center}

\subsection{Background estimation}
\label{sec:con}
The residual background events are from various sources, such as the ISR
production of $J/\psi$ and $\psi(3686)$, $e^+e^-\to (\gamma)\gamma\gamma$,
$e^+e^-\to\psi(3770) \to D\bar{D}$, $e^+e^-\to\psi(3770) \to$ non-$D\bar{D}$,
and $e^+e^- \to$ continuum processes. These background contributions are
estimated by analyzing the corresponding background MC events. The
contamination rate for the candidate Bhabha events is calculated as
$\eta=N^{\rm MC}_{\rm bkg}/(N^{\rm MC}_{\rm BB}+N^{\rm MC}_{\rm bkg})$, where
$N^{\rm MC}_{\rm BB}$ and $N^{\rm MC}_{\rm bkg}$ are the numbers of Bhabha and
background MC events that satisfy the selection criteria, respectively. These
numbers are normalized according to individual cross sections. The resulting
contamination rate is $\eta= 3 \times 10^{-4}$.

\subsection{Detection efficiency and cross section}
\label{sec:eff}
To estimate the detection efficiency for the Bhabha events, an
$e^+e^-\to(\gamma)e^+e^-$ signal MC sample is generated for each run, using the
calibrated $E_{\rm cm}$, with the Babayaga@NLO
generator~\cite{babayaganew, Balossini:2006wc, Balossini:2008xr, theoretical}.
The MC samples are generated within the range of  $|\cos\theta|<0.83$, where
$\theta$ represents the polar angle of the final state $e^+$ and $e^-$. By
applying the same selection criteria as used in the data analysis to these MC
samples, the efficiency is calculated as the ratio of the number of selected
signal MC events to the generated number of events. The cross section for each
run is also estimated using Babayaga@NLO, taking into account the calibrated
$E_{\rm cm}$ of each run. A total of 2~billion signal MC events are generated,
which corresponds to approximately the same size as the data. This large sample
size allows us to neglect the uncertainty arising from the statistics of the
signal MC sample. The detection efficiency and cross section within
$|\cos\theta|<0.83$ at 3.773~GeV are determined to be 61.09\% and 147.47~nb,
respectively, with fluctuations on the order of $0.1\%$ for different runs due
to variations in $E_{\rm cm}$.

\subsection{Integrated luminosities}
The number of observed Bhabha events for each run $N_i^{\rm obs}$ is determined
by counting the events that satisfy the selection criteria outlined in
Sec.~\ref{sec:selection}. In total, $450.97\times10^{6}$, $736.88\times10^{6}$,
and $379.45\times10^{6}$ events are obtained for DATA~I, DATA~II, and DATA~III,
respectively. The trigger efficiency $\epsilon^{\rm trig}$ for collecting
$e^+e^-\to(\gamma)e^+e^-$ events has been measured to be 100\% with a
statistical uncertainty of less than 0.1\%~\cite{Berger:2010my}.

Inserting the number of observed Bhabha events, the trigger efficiency, the
contamination rate~(Sec.~\ref{sec:con}), the detection efficiency, and the
cross sections within $|\cos\theta|<0.83$~(Sec~\ref{sec:eff}) into
Eq.~(\ref{eq:cal_lum}), the integrated luminosities are determined to be
$4.995 \pm 0.019$~fb$^{-1}$ for DATA~I, $8.157 \pm 0.031$~fb$^{-1}$ for
DATA~II, and $4.191 \pm 0.016$~fb$^{-1}$ for DATA~III, where the statistical
uncertainties are  negligible and the systematic uncertainties will be
discussed in the next subsection.

\subsection{Systematic uncertainty}
The systematic uncertainty arising from the MDC information, which includes the
MDC tracking efficiency and the momentum requirement, is determined to be
0.29\% by comparing the integrated luminosities measured with and without the
MDC information. The distribution of $\delta \phi$ without the MDC information
applied is shown in Fig.~\ref{fig:delphi_sys}, and the efficiency increases by
about 5\%. The dominant background is $e^+e^-\to (\gamma)\gamma\gamma$, which
is at the 0.15\% level. The systematic uncertainty associated with the
$E_{\rm EMC}$ requirements is estimated by varying the requirement from 1.0 to
0.9 or 1.1~GeV. This variation results in a change of 0.16\% in the luminosity
for both tracks. To estimate the systematic uncertainty caused by the
$\cos\theta$ requirement, the integrated luminosity is determined with
$|\cos \theta |<0.75$ or 0.70, and the difference from the standard selection
of $|\cos \theta |<0.80$ is found to be 0.13\% for both tracks, which is
assigned as the corresponding systematic uncertainty. The uncertainty due to
the $\delta \phi$ signal region selection is estimated to be 0.02\% by
comparing the integrated luminosities obtained by changing the lower limit
from $5^{\circ}$ to $0^{\circ}$, or the higher limit from $40^{\circ}$ to
$20^{\circ}$ or $30^{\circ}$. The uncertainty arising from the trigger
efficiency~\cite{Berger:2010my} and the theoretically calculated cross
section using the Babayaga@NLO generator~\cite{Balossini:2008xr, bbyg,bbygu}
are both 0.1\%. The uncertainties from the MC statistics and the contamination
rate are negligible.

\begin{center}
  \includegraphics[width=7cm]{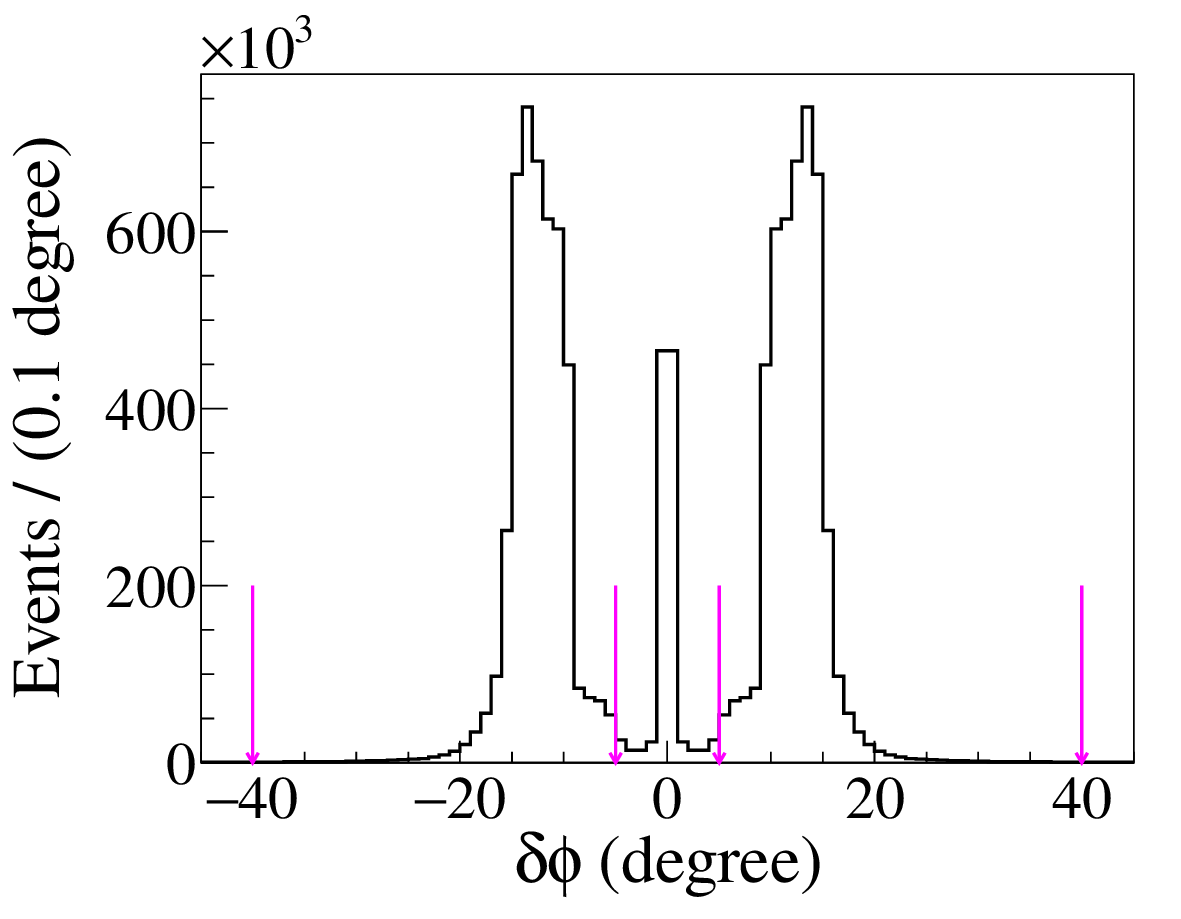}
  \figcaption{
    The distribution of $\delta \phi$ between the selected $e^+$ and $e^-$
    tracks without MDC information from a subset of the data sample.
  }
  \label{fig:delphi_sys}
\end{center}

All contributions to the systematic uncertainty are summarized in
Table~\ref{Tab:err_lum}. The total systematic uncertainty is obtained by
adding all individual contributions in quadrature.
\section{Summary}
By analyzing large angle Bhabha scattering events, we have measured  the
integrated luminosity of the $e^+e^-$ collision data collected at
$E_{\rm cm} = 3.773$~GeV from the years 2021 to 2024 with the BESIII detector
at the BEPCII collider. These are crucial normalization factors for
experimental studies of the production and decays of the $\psi(3770)$ and $D$
mesons. Using the cross sections of
$\sigma(e^+ e^-\to D^0\bar{D}^0)=(3.615 \pm 0.010_{\rm stat} \pm 0.038_{\rm syst})$~nb
and
$\sigma(e^+e^-\to D^+D^-)=(2.830 \pm 0.011_{\rm stat} \pm 0.026_{\rm syst})$~nb~\cite{BESIII:2018iev},
one can obtain the numbers of $D^0\bar D^0$ and $D^+D^-$ pairs in data. All the
numerical results are summarized in Table~\ref{tab:lum}. Along with the
$(2.932\pm0.014)$~fb$^{-1}$ data collected from 2010 to
2011~\cite{BESIII:2015equ, Ablikim:2013ntc}, BESIII has accumulated data sets
with a total integrated luminosity of $(20.275\pm0.077)$~fb$^{-1}$ at
$E_{\rm cm} = 3.773$~GeV. These results offer fundamental inputs for physics
analyses based on these data samples~\cite{Ke:2023qzc, Li:2021iwf}.
\begin{center}
\tabcaption{The relative systematic uncertainties in the luminosity determination.} \footnotesize
\begin{tabular*}{80mm}{l@{\extracolsep{\fill}}ccc}
\toprule Source & Uncertainty~($\%$)  \\
\hline
MDC information & 0.29\\
$E_{\rm EMC}$   & 0.16\\
$\cos\theta$    & 0.13\\
$\delta\phi$    & 0.02\\
Trigger         & 0.10\\
Generator       & 0.10\\
\hline
Total           & 0.38\\
\hline
\bottomrule \label{Tab:err_lum}
\end{tabular*}
\end{center}

\begin{table*}[htp!]
  \centering
  \tabcaption{Numerical results for DATA~I, DATA~II, and DATA~III
    along with data collected from 2010 to 2011~(DATA~2010)~\cite{BESIII:2015equ},
    the number of observed
    events~($N_{\rm obs}$), the integrated luminosity~($\mathcal{L}$), and the numbers of
    $D^0\bar D^0$~($N_{D^0\bar D^0}$) and $D^+D^-$~($N_{D^+D^-}$).} \footnotesize  
  \begin{tabular}{lcccc}
    \toprule
    \hline
    Sample   & $N_{\rm obs}$~($10^{6}$) & $\mathcal{L}$~(fb$^{-1}$) & $N_{D^0\bar D^0}$~($10^{6}$) & $N_{D^+D^-}$~($10^{6}$)\\ 
    \hline
    DATA~I   & 450.97                   & $4.995 \pm 0.019$         & $18.06 \pm 0.21$            & $14.14 \pm 0.16$ \\
    DATA~II  & 736.88                   & $8.157 \pm 0.031$         & $29.49 \pm 0.34$            & $23.08 \pm 0.25$ \\
    DATA~III & 379.45                   & $4.191 \pm 0.016$         & $15.15 \pm 0.18$            & $11.86 \pm 0.13$ \\ 
    DATA~2010~\cite{BESIII:2015equ} & 283.95 & $2.932 \pm 0.014$    & $10.60 \pm 0.13$            & $8.30  \pm 0.09$ \\
    \hline
    TOTAL     & 1851.25                 & $20.275\pm 0.077$         & $73.29 \pm 0.84$            & $57.38 \pm 0.61$ \\
    \hline
    \bottomrule\label{tab:lum}
  \end{tabular}
\end{table*}

\end{multicols}
\vspace{-1mm}
\centerline{\rule{80mm}{0.1pt}}
\vspace{2mm}

\begin{multicols}{2}
  
\end{multicols}

\clearpage

\end{document}